\documentclass[letter]{aa}
\usepackage{graphicx}
\usepackage{epsfig}
\usepackage{latexsym}
\usepackage{txfonts}
\usepackage{natbib}
\usepackage{epstopdf}
\begin{document}
\renewcommand{\labelitemi}{-}
%title & authorsreplac
\title{Height structure of X-ray, EUV and white-light emission in a solar flare}
\author{Marina Battaglia
  \and Eduard P. Kontar }
\institute{School of Physics \& Astronomy, University of Glasgow, Glasgow G12 8QQ, UK}
\offprints{M. Battaglia, \email{marina.battaglia@glasgow.ac.uk}}
\date{Received /Accepted}
%abstract
\abstract
%context
{The bulk of solar flare emission originates from very compact sources located in the lower solar atmosphere and seen in various wavelength ranges: near optical, UV, EUV,  soft and hard X-rays, and gamma-ray emission, yet very few spatially resolved imaging observations to determine the structure
of these compact regions exist.}
%aims
{We investigate the above-the-photosphere heights of hard X-ray (HXR), EUV and white-light  (6173 $\AA$) continuum sources in the low atmosphere
and the corresponding densities at these heights. Considering collisional transport of solar energetic electrons we also determine where and how much energy is deposited and compare these values with the emissions observed in HXR, EUV and continuum.}
%methods
{Simultaneous EUV/continuum images from AIA/HMI on-board SDO and HXR RHESSI images are compared
to study a well observed gamma-ray limb flare. Using RHESSI X-ray visibilities we determine the height of the HXR sources as a function of energy above the photosphere. Co-aligning  AIA/SDO and HMI/SDO images with RHESSI we infer, for the first time, the heights and characteristic densities of HXR,
EUV and continuum (white-light) sources in the flaring footpoint of the loop.}
%results
{35-100 keV HXR sources are found at heights between 1.7 and 0.8 Mm above the photosphere, below the 6173 $\AA$ continuum emission which appears at heights $1.5-3$ Mm, and the peak of EUV emission originating near 3 Mm. }
%conclusions
{The EUV emission locations are consistent with energy deposition from low energy electrons of $\sim 12$~keV
occurring in the top layers of the fully ionized chromosphere/low corona and not by $\gtrsim 20$ keV electrons that produce HXR footpoints
in the lower neutral chromosphere. The maximum of white-light continuum emission appears between the HXR and EUV emission,
presumably in the transition between ionized and neutral atmospheres suggesting free-bound and free-free continuum emission.
We note that the energy deposited by low energy
electrons is sufficient to explain the energetics of optical and UV emissions.}
\titlerunning{Multiwavelength source heights}
\authorrunning{M. Battaglia \& E. P. Kontar}
\maketitle

%\documentclass[onecolumn ]{emulateapj}
%\usepackage{graphicx}
%\usepackage{epsfig}
%\usepackage{latexsym}
%\usepackage{txfonts}
%\usepackage{natbib}
%\begin{document}
%\title{Niceflare}
%
%\author{M. Battaglia and E. P. Kontar}
%\affil{SUPA, School of Physics \& Astronomy, University of Glasgow, G12 8QQ, UK}
%\email{e-mail: Marina.Battaglia@glasgow.ac.uk}
%

%\begin{abstract}
%
%\end{abstract}
%
%\keywords{Sun: flares -- Sun: X-rays, $\gamma$-rays -- Sun: Chromosphere -- Acceleration of
%particles}
%

% Introduction
%-------------------------------------------------------
\section{Introduction} \label{Introduction}
Solar flare emission is now observed virtually over the whole electromagnetic
spectrum from radio frequencies as low as $10-100$~MHz up to photon energies
as high as a few hundred MeV. The bulk of this emission
originates from the chromosphere, a small part of the solar atmosphere which is narrow in height
and much smaller than the size of a flaring region. Being only about $<3$ Mm
($<4''$) across, the chromosphere presents a formidable observational challenge
in both angular and temporal resolution especially for transient phenomena like solar flares.
During solar flares large numbers of energetic electrons are accelerated in the
solar corona and effectively release their energy in the dense chromosphere.
These energetic particles are believed to be responsible (although not always directly) for HXR, EUV, continuum "white-light" (WL),
infrared and radio emissions. On the basis of temporal correlation, the interconnection between these emissions was early
realized \citep{Na70,Sv70}. Due to the lack of height-resolving observations spectroscopic data have been mostly
used to identify the structure of the emitting region \citep[e.g.][]{Ca74}. Therefore, the actual mechanisms of WL
emission and the relation to HXR emitting electrons is still subject of debate \citep{Ne89,Ne93,2000SoPh..194..305S,Mat03,Po10,Wa10,2011A&A...530A..84K}.
It is not even clear whether the emission is optically thick or thin or both. Thus the characteristic heights of continuum WL
emission in various models are placed from the photosphere to the upper chromosphere.  Observations supporting both \citep{1995A&AS..110...99F, Xu06},
a photospheric origin \citep[e.g.][]{Bo85,1999ApJ...512..454D,Ch06} presumably due to radiative back-warming,
and generation of WL emission in the higher temperature regions of the upper chromosphere \citep[e.g.][]{Ma74, Hu72}
have been made.  The major observational challenge here is to have simultaneous HXR, UV and WL images
with sufficiently high angular resolution. The relative horizontal positions of HXR and WL flares have been investigated in detail and,
although generally been found to agree \citep[e.g.][]{Kr11, Fl07,Hu06,Me03}, the uncertainty in the relative position
was large due to limited absolute pointing accuracy of partial solar disk observations such as {\it TRACE}.
Thus it was not possible to quantify the relative height of the sources above the photosphere in previous studies.
With RHESSI \citep{Li02} data, the height structure of HXR sources has become accessible for detailed
observational studies \citep{As02, Ko08,Ko10, Ba11a, Sa10}. These observations demonstrated that
the higher energy HXR footpoints originate from the chromosphere from progressively lower heights,
in good agreement with collisional downward propagation of energetic electrons \citep[see][for details]{Br02}.
The newly available SDO continuum images from HMI \citep{Wa11} and AIA EUV \citep{Le11}
observations allow to scrutinize the spatial structure and more importantly the height structure of various emissions
associated with HXR footpoints.
In this paper we present the first spatially resolved simultaneous observations of HXR, EUV and WL emission from a footpoint
of a solar limb flare observed with RHESSI, SDO/AIA  and HMI. Using the recently developed hard X-ray visibility analysis
technique \citep{Hur02,Sc07} we determine the characteristic heights for the different types of emission using the method
presented by \citet{Ba11a, Koet10, Ko08}. The results are consistent with the EUV and continuum WL being produced
above the HXR footpoints by lower energy electrons ($\sim 12$ keV)  and hence suggest not a photospheric
but upper chromospheric origin of WL emission.
\section{Height and density measurements in HXR and EUV}\label{srheight}
\begin{figure}
\centering
%\resizebox{\hsize}{!}{\includegraphics{paperplot_may/aia_rhessi_light_curves.pdf}}
\includegraphics[width=7.2cm]{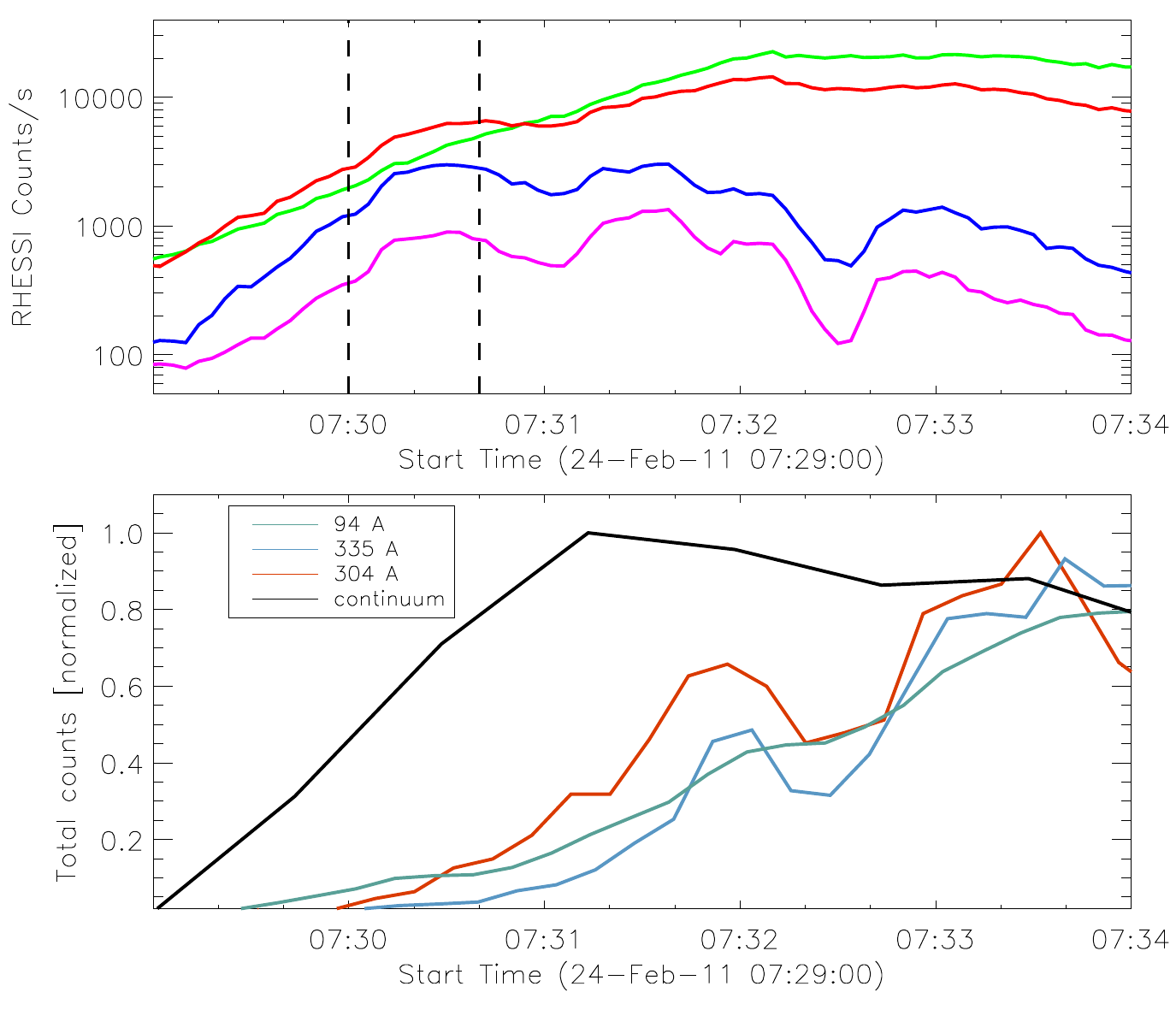}
\caption {Top: RHESSI lightcurves of the main flare phase in different energy bands: 6-12 keV (green), 12-25 keV (red), 25-50 keV (blue), 50-100 keV (purple). Bottom: Normalized AIA and HMI emission from location of the southern HXR footpoint as function of time in several wavelengths (see legend). }
\label{lightcurves}
\end{figure}
The GOES M3.5 limb flare happened on February 24th 2011 with three well-pronounced main HXR (above $\sim 30$ keV) peaks
 between 07:29 and 07:33 UT and an associated filament eruption to the south of the flaring site. Figure \ref{lightcurves} (top) shows
the RHESSI lightcurves in different energy bands of the thermal and non-thermal emission. The attenuator state was 1 during the main
phase of the flare and the live-time was better than 90\% during the whole period of the observations, hence making the effects of pulse
pile-up negligible.  The flare appeared in active region AR11163 and is well observed in enhanced EUV emission visible in all AIA filters,
and also showed 6173 $\AA$ continuum emission observed by HMI. The continuum emission (with a maximum enhancement of $\sim15$\%
above the photospheric background) shows two footpoints and follows the X-ray time profile.
Unfortunately, AIA images became saturated in most wavelengths before the soft X-ray peak (2nd HXR peak) of the flare.
Because of this saturation, we focus the analysis on the time of the first HXR peak.
During the time around 07:30:10 UT, AIA was not saturated in 94 $\AA$, 193 $\AA$, 304 $\AA$ and 335 $\AA$ filters near the location of the southern footpoint.
%\begin{figure*} %[!]
%\begin{minipage}[l]{0.5\textwidth}
%%\includegraphics[width=9.5cm,bb=00 00 400 400]{wl_rhessi_test.png}
%\includegraphics[width=9.5cm]{wl_flare_rhessi171_new.pdf}
%\end{minipage}
%\begin{minipage}[r]{0.5\textwidth}
%\includegraphics[height=8.4cm]{aia_difference_images_4p.pdf}
%\end{minipage}
%\caption{Left: AIA 171 $\AA$ image overlaid with RHESSI soft (red) and hard X-ray emission (blue). Right: AIA difference images in 3 wavelengths and HMI image (bottom right). Contours at 50, 70 and 90 \% of the maximum emission in RHESSI CLEAN images between 07:30:00 and 07:30:40 UT
%at 6-12 keV (red) and 25-50 keV (blue) are overlaid}
%\label{aiadiff}
%\end{figure*}
\begin{figure} %[!]
\centering
\includegraphics[width=7.6cm]{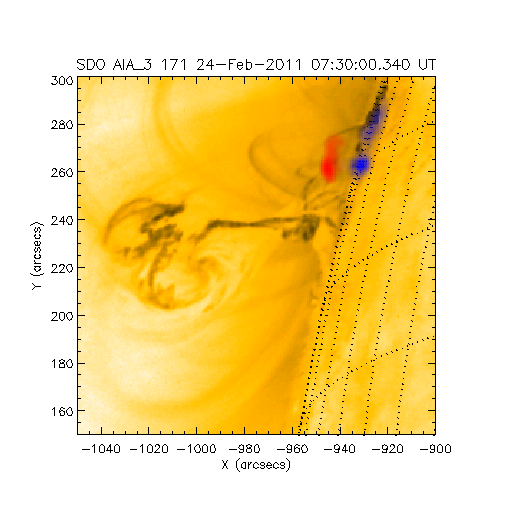}
\includegraphics[width=6.7cm]{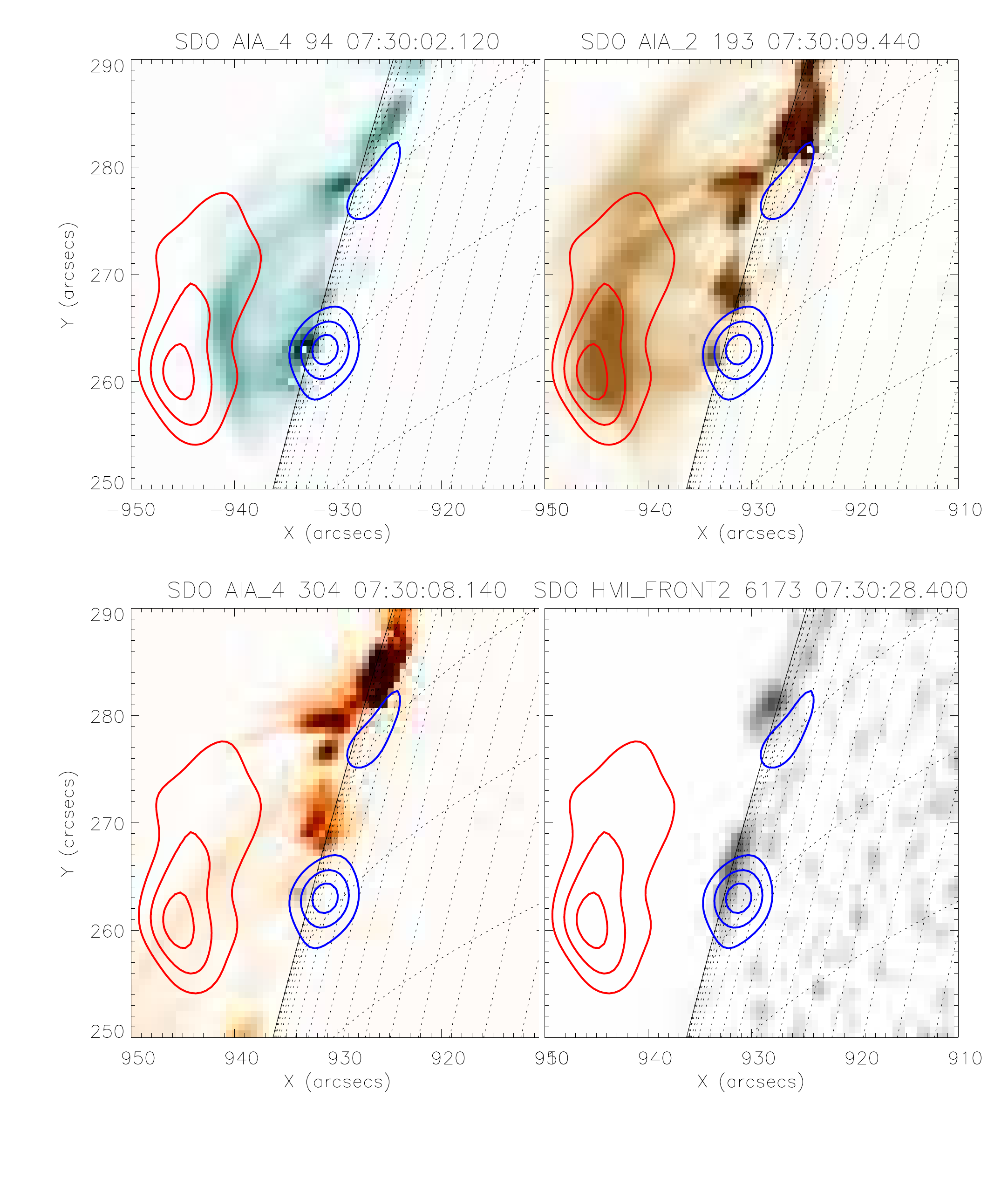}
\caption{Top: AIA 171 $\AA$ image overlaid with RHESSI soft (red) and hard X-ray emission (blue). Bottom: AIA difference images in 3 wavelengths and HMI image (bottom right). Contours at 50, 70 and 90 \% of the maximum emission in RHESSI CLEAN images between 07:30:00 and 07:30:40 UT
at 6-12 keV (red) and 25-50 keV (blue) are overlaid.}
\label{aiadiff}
\end{figure}
To study the flare related enhancement in EUV and continuum emissions the image from one minute before the analyzed time was subtracted.
 Figure \ref{aiadiff} displays difference images in AIA 94, 193 and 304 $\AA$ wavelengths and the HMI
 continuum\footnote{Animated images of the flare are available online: http://www.astro.gla.ac.uk/users/mbattaglia/20110224\_online\_material/}.
For the RHESSI analysis a 40 second time interval at the first peak of the flare from 07:30:00 UT was chosen (see Fig. \ref{lightcurves}).
The contours in Fig. \ref{aiadiff} indicate the RHESSI emission between 07:30:00 and 07:30:40 UT at 50, 70 and 90 \%
of the maximum in CLEAN images \citep{Hur02}. The images show a coronal (SXR) source above the limb
and two footpoints (HXR) sources (north and south). During the first peak the southern footpoint source was much more
 intense than the northern footpoint. Thus we focused the height-analysis on the southern footpoint.
We note that in the subsequent peaks the northern footpoint reached the same intensity as the southern footpoint.
The RHESSI spectrum suggests emission well beyond 200 keV into the $\gamma$-ray range. During the analyzed time interval,
the HXR spectrum is well fitted with a thermal component and a single thick-target power-law with spectral index $\delta=3.9$.
The flare was also observed by Fermi/GBM \citep{Mee09} and the corresponding spectrum from FERMI data
was found virtually identical to the RHESSI spectrum. The time evolution of the EUV emission at the position
of the southern HXR footpoint can be seen in Fig. \ref{lightcurves} (bottom). The total flux in a region of 5 by 10 arcsec
extent  around the position of the southern HXR footpoint is shown.

\begin{figure}[]
\centering
%\resizebox{\hsize}{!}{\includegraphics{paperplot_may/aia_height_cont_edep.pdf}}
\includegraphics[width=8.2cm]{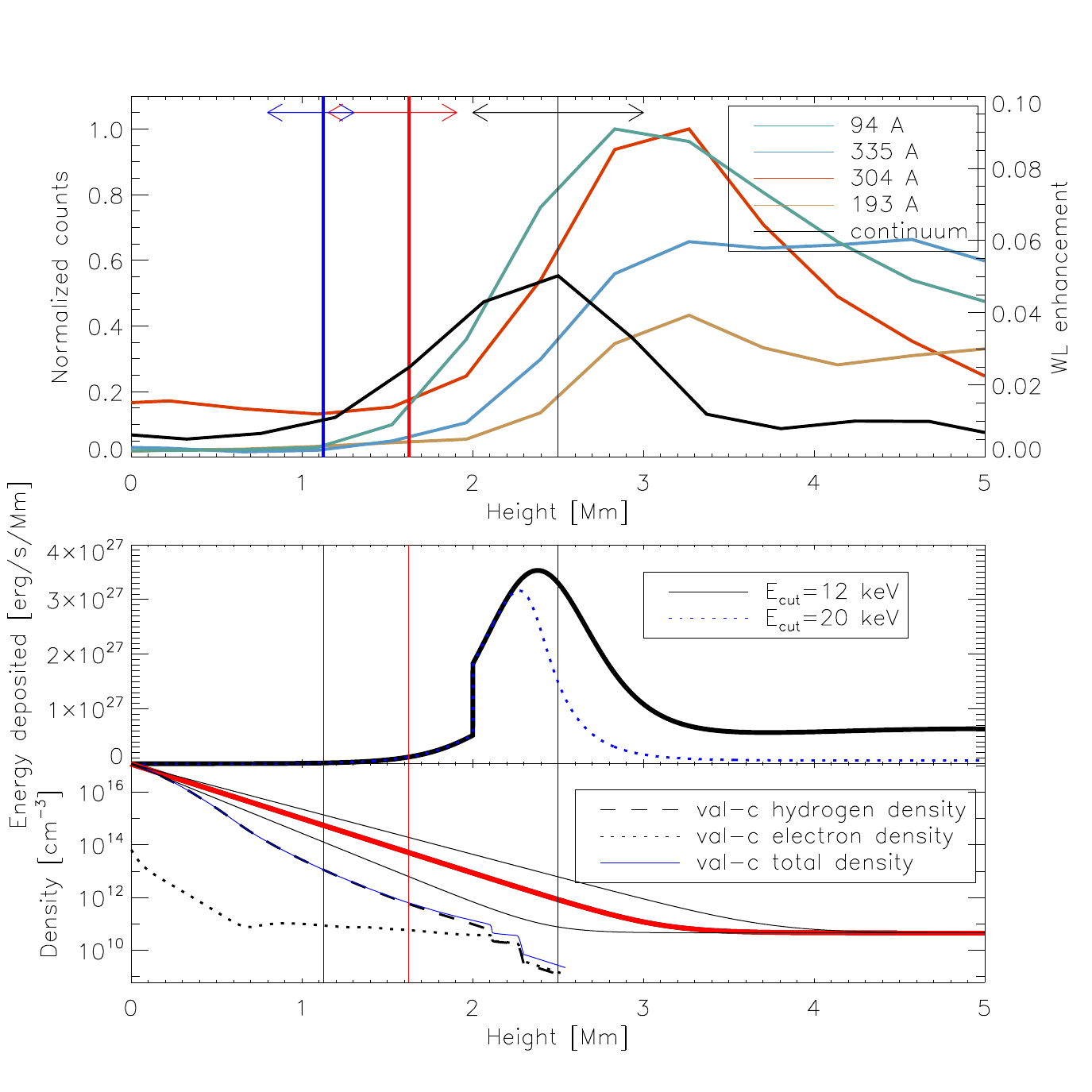}
\caption {Top: Height of HXR emission at the lowest (red vertical lines) and highest (blue vertical lines) imaged energies, EUV emission at different wavelengths (see legend) and WL relative enhancement where the black line indicates the height of maximum enhancement. Arrows indicate the corresponding uncertainties. Middle: Energy deposition rate as a function of height for two different cases of electron cutoff energy. Bottom: Fitted density model (thick solid line). The thin solid lines give the lower and higher limits of the density due to the uncertainty in the fit of the scale height. The model from \citet{Ver81} is given for comparison.}
\label{aia_height}
\end{figure}

%\section{Height and density measurements in HXR and EUV} \label{srheight}
Because of the limb location of the flare the height analysis is possible. However,
SDO observations alone cannot infer the heights of the sources in different EUV wavelengths or
for white-light emission. Here, utilizing RHESSI observations, we find the density structure of the chromosphere
and importantly the reference height (projected radius of the point at the photospheric level beneath the HXR sources)
as described by \citet{Ba11a} and \citet{Koet10,Ko08}. Following the method described by those authors
we use visibility forward fitting to find the position of the southern footpoint as a function of HXR energy
in several energy bands (30-40 keV, 40-55 keV, 55-80 keV, 80-130 keV). The energy bands were chosen
large enough to guarantee good count statistics for reliable fits. Both the southern and the northern footpoint
are fitted with circular Gaussians. This assures that the emission of the northern footpoint is properly accounted for,
even if it is not intense enough to result in fit parameters with acceptable errors.
The flare morphology was such that the radial direction (direction of electron propagation along the loop)
can reasonably be seen as along the x-direction in the images (see Fig. \ref{aiadiff}). An exponential density profile
(hydrostatic atmosphere at lower heights), $n(h)=n_{cs}+n_{ph}\exp(-(h-r_{ref})/h_0)$, was fitted to the positions
found by visibility forward fitting, where $n_{cs}=4.6\times 10^{10}$ $\mathrm{cm^{-3}}$
is the (constant) loop density, $n_{ph}=1.16\times 10^{17}$ $\mathrm{cm^{-3}}$ the photospheric
density \citep[][]{Ver81} which corresponds to $\tau_{500 nm}=1$, and $h$ is the height above the photosphere
\footnote{In this work we use a fit to the full expression for the HXR flux maximum and not the simplified
version given by Eq. 7 of \citet{Br02}. The latter is likely to underestimate the densities in a sharply changing
chromosphere. See also \citet{Ba11a, Koet10, Ko08}}. The loop density $n_{cs}$ was assumed to be the same as the coronal source density determined from the emission measure given by the spectral fit
and the size of the coronal source in RHESSI 6-12 keV images.
Fitting the HXR energy-position relation with the above density model we find a scale height of  $h_0=211 \pm 46$ km and a reference height for the radial
distance between the solar disk center and the point under the footpoints at the photosphere of $r_{ref}=929.4\pm0.3$ arcsec.
The photospheric reference distance $r_{ref}$ is used to find the absolute height of the EUV and WL sources
above the photosphere. Using the HXR source size as the size of the magnetic flux tube,  we selected a stripe along the x-direction
of 5 arcsec width in the y-direction over the position of the southern RHESSI footpoint.
At each horizontal distance $x$, the AIA and HMI counts were summed over the y-direction.
The height of the emission is simply the $x$-position minus $r_{ref}$ as found above.
This is shown in Fig. \ref{aia_height}.  Proper co-alignment of the images is crucial for this study.
The RHESSI disk center is known to better than 0.2 arcsec accuracy. Using SDO full disk images the disk center
of AIA and HMI can be determined. This position of solar-disk center is found to deviate by only -0.39/0.67 arcsec in x/y
direction that of RHESSI, which, although small, were taken into account for the co-alignment.
The larger uncertainty is related with the roll-angle of SDO images (rotation around the disk center). To investigate the effect,
the HMI image was rotated by $0.1$ degree relative to Sun center. This uncertainty translates to mostly $y$-direction
uncertainty of $<1.6$ arcseconds and $<0.5$ for the $x$-direction for the limb event under study.
For the height measurements of the EUV and HMI emissions relative to the RHESSI emission,
we have taken this uncertainty into account and shown in Fig. \ref{aia_height}.
We find that the RHESSI sources at 35 - 100 keV originate from heights between 1.7 and 0.8 Mm.
The WL emission peaks at 2.5 Mm, which is $\sim1$ Mm above the 30 keV HXR emission,
suggesting it originates from the upper chromosphere.
%\subsection{Energy deposition}
Using the density model found above and a power-law electron 
injection rate $F_0(E_0)=F_{norm}E_0^{-\delta}$ [electrons s$^{-1}$ keV$^{-1}$] 
found from the spatially integrated RHESSI X-ray spectrum, we can compute the energy deposition 
rate [ergs s$^{-1}$ Mm$^{-1}$]  as a function of height \citep{Br73a},
\begin{equation} \label{eqdudx}
\frac{dU}{dx}=K[\Lambda _{eH}n_H(h)+\Lambda _{ee}n_e(h)]\int_{E}\frac{F(E,h)}{E}\mathrm{d}E
\end{equation}
where $F(E,h)$ is the electron rate spectrum \citep{Br71}, $K=2\pi e^4 $ with $\Lambda_{eH}$  and $\Lambda_{ee}$ the Coulomb logarithms for electron-neutral and electron-electron collisions, and  $n_H(h)$ and $n_e(h)$  are the hydrogen 
and electron number densities respectively.
The resulting energy deposition rate as a function of height given by Equation (\ref{eqdudx}) is shown in Fig. \ref{aia_height} where we assumed a step change in ionization from fully ionized to neutral at a height of 2 Mm.
As expected, the deposition rate is dependent on the low energy cut-off for the injected electron spectrum
$F_0(E_0)$. From the thick target fit to the RHESSI spatially integrated HXR spectrum 
the cut-off energy is around $E_c=12$ keV and the spectral index $\delta$=3.9. 
This leads to a total power of $\sim 2.7\times 10^{27} \;\mathrm{erg s^{-1}}$. However, 
the cut-off is generally unknown as the low energy part of X-ray spectrum is dominated by thermal emission,
so lower cut-off values are possible and likely \citep[e.g.][]{Emg03,Ha09}.
However, even with $E_c=12$ keV, the energy deposited appears much larger than emitted in other wavelengths.
For the southern footpoint, an enhancement of $\sim 5$\% over the total solar irradiance of $3.4 \times 10^{26}\mathrm{erg\,s^{-1}arcsec^{-2}}$ 
over a source area of $\sim 20$ arcsec$^2$ leads to a total power emitted of $3.4\times10^{26}\mathrm{erg\,s^{-1}}$. 
The energy in the non-thermal electrons is therefore up to an order of magnitude larger than what is needed to power the WL emission, which often constitutes the bulk of flare emitted energy \citep{Ne89}. We note that the WL energetics obtained using 6173 $\AA$ continuum is found to be lower than using TRACE wavelengths near the 1700 $\AA$ continuum, and assuming black body emission to deal with the TRACE response \citep[e.g.][]{Fl07}, but close to the values reported by \citet{Ne89} based on spectroscopy above 2500 $\AA$. Indeed, various optical bands show different levels of enhancement during flares \citep{2011A&A...530A..84K}. In addition, the maximum of 6173 $\AA$ continuum is reached later in the flare, after the analyzed time interval.

\section{Discussion and Conclusions}
Our observations suggest that the WL emission originates from the upper chromosphere or lower transition 
region with a density of between  $10^{11}\mathrm{cm^{-3}}$ and $10^{13}\mathrm{cm^{-3}}$. 
These upper and lower limits of the density are due to the uncertainties in the fit of the density scale height. 
Figure \ref{aia_height} (bottom) displays the fitted density model, as well as the upper and lower 
limits for the density. The arrows in Fig. \ref{aia_height} (top) illustrate the uncertainty in the RHESSI source 
heights resulting from the uncertainty in the density scale height.
The peak of WL emission is found about 1 Mm higher than the position of the lowest observed RHESSI energy 
range (30 - 40 keV). Note that despite all the uncertainties in the height due to the density fit or due to relative pointing error,
the WL source is still above the 30-40 keV RHESSI sources. Due to the presence of the SXR coronal emission in the X-ray spectrum, 
the non-thermal HXR spectrum below $\approx$ 20 keV is not known. However, extrapolating the height function found 
in Section \ref{srheight} to lower energies, a source at 12 keV would be seen at about 2 Mm. 
This coincides with the height of the maximum energy deposition as found from Eq. \ref{eqdudx} 
for cut-off energies of 12 keV (Fig. \ref{aiadiff}). The observed position of the maximum of the WL emission 
would then hint towards energy deposition by the low energy part of the non-thermal electron distribution, 
contrary to the often made assumption of electron cut-off energies in the range of 20 keV. 
This also suggest that WL and HXR sources above 30 keV should be spatially separated by $<$ 1 Mm with decreasing separation in events towards the disk center. However, we note that the height of the maximum energy deposition, especially in the case of low initial cut-off energies, is influenced by the loop density 
and the length of the loop. Assuming that the acceleration of the particles happens in the center of what is observed 
as the coronal source Fig. \ref{aiadiff} indicates a precipitation
distance of about 13 arcsec from the site of acceleration to the footpoints. This value was used for the computation 
of the energy deposition rate. In this case, the maximum energy deposition of electrons with energy $\lesssim 12$ keV will be near the top of the loop. Should the loop density be lower or the acceleration region extended,
the lower energy electrons would be depositing their energy deeper.
The other interesting aspect of this flare is that HXR emission above $20$~keV is well fitted by a single 
power-law. Non-uniform ionization of plasma produces a break at the energy 
corresponding to the column depth of the transition region, which is observed in other 
flares \citep{2002SoPh..210..419K,2011ApJ...731..106S}. The absence of such break points 
to the transition region above the stopping depth of 20 keV electrons again is consistent with the 
height measurements.
The observation of strong emission from heights  $\ge 2$ Mm suggests that a substantial part of the WL continuum is formed in an optically thin 
or finite optically thick region with ionized or partially ionized plasma in the upper chromosphere favoring free-bound 
and free-free emission of the $6173 \AA$ continuum.
This is additionally supported by the height of the AIA emission at 94, 335 and 193 $\AA$ (representing temperatures $>10^6$ K) 
originating from around 3 Mm and above the WL emission peak. Finally, it is likely that the WL is powered by lower energy
electrons $\sim 12$~keV, which do not penetrate deep into the solar chromosphere.

\mdseries
\begin{acknowledgements}
The authors thank the anonymous referee for helpful comments and suggestions. This work is supported by the Leverhulme Trust (M.B., E.P.K.),
STFC rolling grant (E.P.K.). Financial support by the European Commission
through the FP7 HESPE network is gratefully acknowledged.
\end{acknowledgements}

\bibliographystyle{aa}
\bibliography{mybib}

\end{document}